# Improvements to the LC Muon Tracking and Identification Software

C. Milsténe, G. Fisk, A. Para
*FNAL, Batavia, IL 60510, USA*

This note summarizes the evolution of the Muon-ID package originally written by R. Markeloff at NIU. The original method used a helical swimmer to extrapolate the tracks from the interaction point and to collect hits in all sub-detectors: the electromagnetic and hadronic calorimeters and muon detector. The package was modified to replace the swimmer by a stepper which does account for both the effects of the magnetic field and for the losses by ionization in the material encountered by the particle. The modified package shows a substantial improvement in the efficiency of muon identification. Further improvement should be reached by accounting for stochastic processes via the utilization of a Kalman filter.

## 1. INTRODUCTION

Muon identification procedure involves a comparison of the charged track trajectory reconstructed in the tracking systems with the location of energy deposits detected in the calorimeters and the muon detectors. This comparison requires that the track trajectory is extrapolated beyond the tracking detectors. The initial extrapolation method, the swimmer, was based on the helical extrapolation of the track. While it is an acceptable approximation for large momentum tracks, it fails at the low momentum end of the spectrum where the energy loss and multiple scattering cause significant deviation of the actual track trajectory from an ideal helix, the discrepancy being amplified by the bending effect in the strong magnetic field.

The Stepper software propagates the charged track through the detector in small steps accounting for dE/dx and $q\vec{v}\times\vec{B}$ effects [1] at each step. It improves the identification and reconstruction efficiency of low energy muons by up to 40%.

The Kalman filter should further improve the results by taking into account multiple scattering, Bremsstrahlung and other stochastic processes. This note describes the general principles of the muon reconstruction package and its implementation.

## 2. THE STEPPER

The stepper starts with a particle at the interaction point (IP) and it computes step-wise the particle trajectory throughout the complete detector. A uniform axial magnetic field $B_z$ is assumed. The momentum components $p_x$ and $p_y$ undergo changes due to the qv×B term whereas the energy loss in material contributes to a reduction of all the components of particle momentum.

### 2.1. The Parametrization

Each component of the momentum changes at each step. There is a momentum change $\delta p_x(B_z)$, $\delta p_y(B_z)$, due to the magnetic field, and one of $\delta p_x(material)=\gamma_x$, $\delta p_y(material)=\gamma_y$, $\delta p_z(material)=\gamma_z$ due to energy loss by ionization in material.
\. In the equations below, q is the charge, $B_z$ the magnetic field, dt(n) the time spent and ds the path length in one step.

The algorithm for determining the particles trajectory including energy loss is given by the equations below. In these equations, q is the charge, $B_z$ the magnetic field, dt(n) the time spent and ds the path length in one step.

$$p_x(n+1) = p_x(n) - 0.3 * q * \frac{p_y}{E(n)} * c_{light} * B_z * \delta t(n) - \gamma_x(n) ;$$

$$p_y(n+1) = p_y(n) + 0.3 * q * \frac{p_x}{E(n)} * c_{light} * B_z * \delta t(n) - \gamma_y(n) ;$$

$$p_z(n+1) = p_z(n) - \gamma_z(n) ;$$

$$\gamma_i(n) = \frac{dE}{d_i} * \frac{E(n)}{|p(n)|} * \frac{p_i(n)}{|p(n)|} * \delta s ; \quad i = x, y, z .$$

Mixed units are used, $p_x$, $p_y$, $p_z$ are in GeV/c, E(n) in GeV, $c_{light}$ =3x10$^8$ m/s, $\delta t$ in seconds, $B_z$ in Tesla

The point ( x(n+1),y(n+1),z(n+1) ) is the position at step n+1, after the momentum change to $p_{x,y,z}$(n+1) at step n.

$$x(n+1) = x(n) + \frac{p_x(n+1)}{E(n+1)} * c_{light} * \delta t(n) ;$$

$$y(n+1) = y(n) + \frac{p_y(n+1)}{E(n+1)} * c_{light} * \delta t(n) ;$$

$$z(n+1) = z(n) + \frac{p_z(n+1)}{E(n+1)} * c_{light} * \delta t(n) .$$

## 2.2. The Muon Candidate

The muon identification algorithm requires a well fitted charged track consistent with the observed energy deposits in EM and HAD calorimeters and at least 12 hits /12 Layers within an extrapolated (θ, φ) road in the muon detector. Calorimetric cells are associated with a track if they are within 2 angular bins in the EM calorimeter and the muon detector or 3 angular bins in the HAD calorimeter, ( Δφ=Δθ= π/840 in EMCal , Δφ=Δθ= π/600 in HADCal, Δφ=Δθ= π/150 in the muon detector). The current version of the muon ID package is restricted to the barrel detector, defined as 0.95<θ<2.2.

## 2.3. Improvement in Track Reconstruction

We have compared the performance of the muon identification algorithm using the stepper with that of the original package of R. Markeloff using the standard swimmer for the track extrapolation to calorimeters and muon chambers. The swimmer approach does not account for the energy loss in the calorimeters, hence it fails for low momentum muons. We have improved its performance for the low momentum tracks by introducing momentum-dependent cuts, (δφ, δθ ~ 1/p), on the distance between the extrapolated track and the calorimeter/muon detector cell. This version of the package is denoted as 'Swimmer + Ad Hoc dE/dx'.

In Figure 1 and Table I are shown the muon–ID efficiency improvement with the evolution of the algorithm for single muons from 3GeV to 50 GeV.

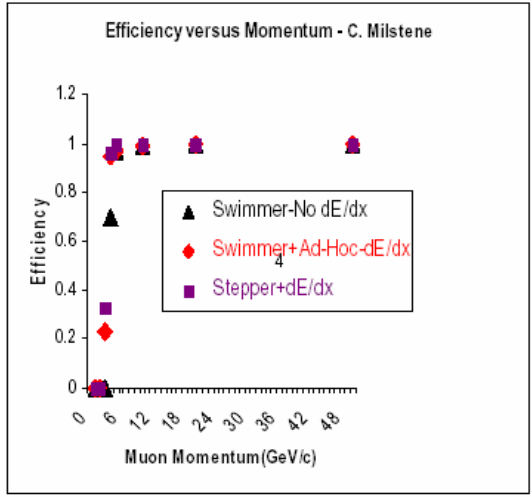

| E(GeV) \ Techn. | 3 | 4 | 5 | 10 |
|---|---|---|---|---|
| No dE/dx | 0.06% | 70% | 97% | 99.% |
| Ad-Hoc dE/dx | 23% | 95% | 97% | 99.% |
| V x B + dE/dx | 33% | 96% | 99% | 100% |

Figure 1& Table I: Muon-ID Efficiency as a function of the momentum

The histogram in Figure 2 shows the angular deviation of the observed hit from the extrapolated track trajectory in φ at different depth (layer number) in the muon detector. Δ(φTrack-φHits) for 20 GeV muons (left) is typically ~one φbin wide whereas Δ(φTrack-φHits) for a 4 GeV muon is ~ 4 times bigger. The remaining inefficiency at low muon momentum is primarily due to the stochastic nature of processes not accounted for by the stepper, e.g. multiple scattering, Beamstrahlung or decays.

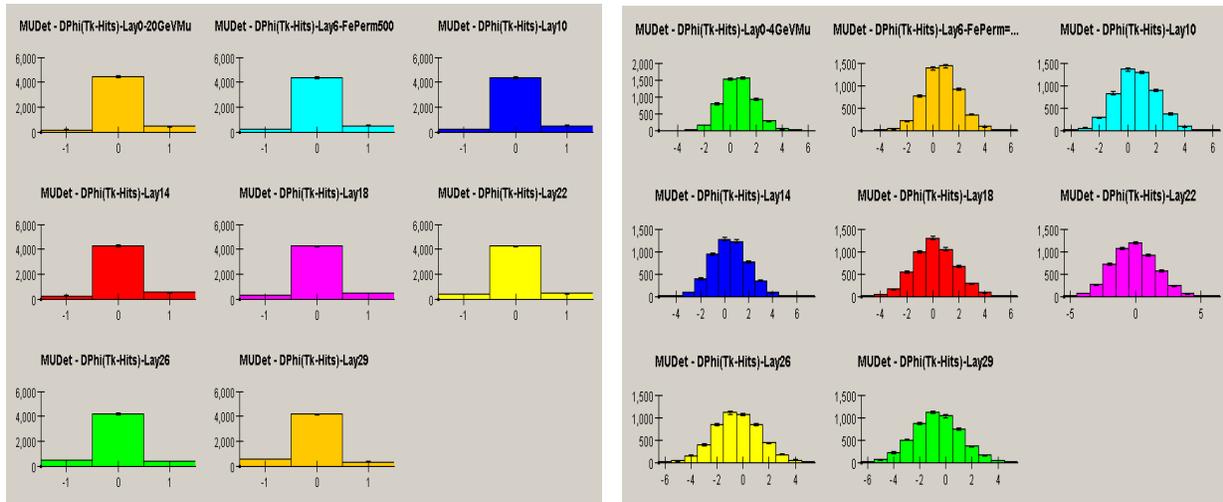

Figure 2: Angular Resolution in φ in MUDet at different radii –left 20 GeV Muon, right 4 GeV Muons

### 2.4. Muon Identification in Jets

Reconstruction and identification of single muons is a relatively easy task. The more difficult, albeit more realistic, case involves muon identification in high energy jets. In such an environment, in addition to the issue of the identification efficiency, we encounter an issue of the purity of the 'muon' sample. Most of particles in jets are hadrons, hence even with small probability of mis-identification of the fake muons may overwhelm the muon candidates sample. To study these issues we have generated a sample of 10,000 b-quark pair events produced in $e^+$ $e^-$ collisions at the center of mass energy up to 500 GeV. Table II shows a breakdown of generated charged particles in this sample. Particles below 3 GeV do not penetrate the muon detector, hence they cannot be identified as muons.

A significant fraction of hadron-induced showers spills over to the muon detectors and is classified as muons by the muon identification algorithm as described earlier. The rate of such fake muon candidates can be reduced by exploiting the fact that the hadronic showers deposit most of their energy in the innermost layers of the hadron calorimeter, as illustrated in Fig. 3. We have augmented the muon ID package by adding a requirement that there are hits along the muon candidate trajectory in the last four layers of the hadron calorimeter. This requirement does not lead to any additional losses of the genuine muons, but it reduces the rates of mis-identified hadrons to those listed in the Table. II. It is worthwhile to notice that 12 of the mis-identified pions and 3 of the kaons are, in fact, decays thus resulting in a genuine muon present in the muon detector.

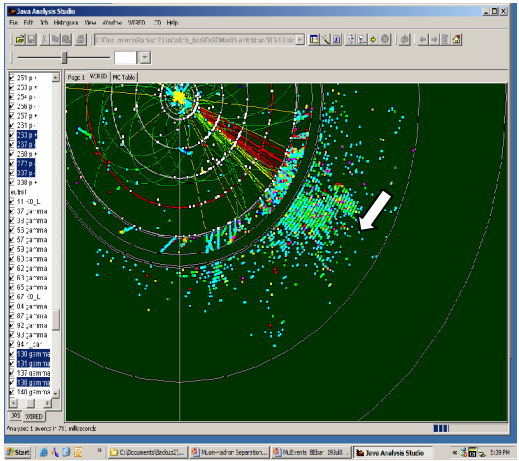

|  | π | K | **Protons** | **Muons** |
|---|---|---|---|---|
| Total Gen. | 55805* | 8310* | 2816* | 1147 |
| Gen.>3GeV | 18666 | 4473 | 1622 | 787 |
| Fraction >3GeV | 34% | 54% | 58% | 69% |
| Recon Tracks>3GeV | 18024 | 4304 | 1614 | 739 |
| Identified as µ | 70 | 41 | 2 | 657 |
| Rejection /ID Efficiency | 1/267 | 1/109 | 1/811 | 83.5% |

Figure 3: Hadronic shower initiated by a jet        Table II: Muon ID and Hadron Rejection in Jets

The average efficiency of the muon reconstruction and identification in jets turns out to be about 83.5%.which is accounted for principally by the abundance of low energy muons. We start from reconstructed tracks by the tracker.

The muon identification package based on the stepper approach shows an improvement over the previous version utilizing the swimmer. The improved version of the swimmer-based algorithm, utilizing an ad-hoc dE/dx implementation, identifies only 76.6% of the muons and the swimmer-based original algorithm alone 65.3% of the muons.

All performance figures quoted above for the efficiency of background rejection and signal identification refer to the number of tracks above 3 GeV. The fake rate probabilities for π's, K's and protons, including punch-through, given by the stepper algorithm are 0.0037, 0.0092 and 0.0012, respectively.

The hadron filtering using the 4 last layers of HADCal suggests that a study of 2 to 4 sensitive planes with a finer grain, located between the coil and the muon detector could be useful

## 3. THE KALMAN FILTER

We expect that further improvement of the muon identification efficiency can be accomplished if the predicted trajectory of the muon could account for actual occurrences of stochastic processes, like multiple scattering. At the same time we expect that background rejection capabilities can be improved if the decay kinks can be detected.

Both of these goals can be achieved, in as much as it is possible, by the application of a Kalman filter technique. Its main advantage over the other methods of track extrapolation stems from its ability to use actual measurements to adjust local track parameters and to offer a better prediction for the track trajectory.

### 3.1. The Principle

We have chosen a 6-dimensional phase space point (x, y ,z, $p_x$, $p_y$, $p_z$) to represent a state vector. The state vector at location k-1, is propagated using a propagation matrix, to location k. The choice of the state vector as the phase point allows the use of the stepper algorithm written in a matrix form as the propagation matrix. Propagation is done in small steps, in general smaller than the thickness of the absorber plates. The multiple scattering in the material is included in the covariance matrix. In the active material, that provides a measurement of the actual track position, the Kalman filter weighting procedure is applied using the estimate at k: xk(-) and the actual measurement at k, $z_k$. At this point, a new state vector $x_k$(+). Is calculated using the weighting procedure which combines the extrapolated and measured track positions. As shown in Figure 4, the changes in vector state accounts for the dE/dx and $B_z$, it is taken into account in the propagation matrix (step 1, 2, 3, 4, 5 ) in the passive material. The multiple scattering is taken care of by a change in the covariant matrix (steps 1, 2, 3, 4, 5).The change in the vector state which accounts for the Kalman Weighting as a result of all the above, takes place when the hit is recorded in the active material (step 6)

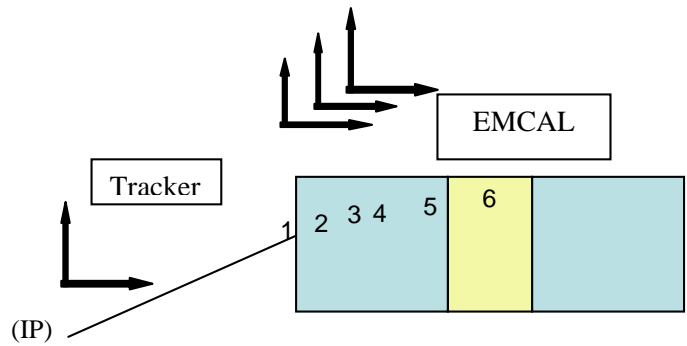

Figure 4: Illustration of the Kalman filter procedure applied in the EMCAL. Five steps over the absorber slabs accumulate the multiple scattering contribution to the covariance matrix; step six in the active material leads to an improvement of the local track parameters.

### 3.2. Code Implementation

The Kalman filter uses the following equations

$$\left\{\begin{array}{l}\vec{x}_{k+1}(-) = \Phi_k \bullet \vec{x}_k(-) \\ P_{k+1}(-) = \Phi_k \bullet P_k(-) \bullet \Phi_k^T + Q_k\end{array}\right\} \quad (1)$$

$$\left\{\begin{array}{l}\vec{x}_k(+) = \vec{x}_k(-) + K_k \bullet [\vec{z}_k - H_k \bullet \vec{x}_k(-)] \\ P_k(+) = [1 - K_k \bullet H_k] \bullet P_k(-)\end{array}\right\} \quad (2)$$

$$K_k = P_k(-) \bullet H_k^T \bullet [H_k \bullet P_k(-) \bullet H_k^T + R_k]^{-1} \quad (3)$$

$$Q_k = |\vec{p}| \bullet \Theta_0 \bullet I$$

Where $Q_k$ is the noise from multiple scattering, $R_k$ = measurement error matrix with dx, dy, dz in the diagonal, $H_k$ = measurement matrix, $\Phi_k$ = propagation matrix, applied in passive material, $x_k(-)$ is the extrapolated vector state $(x,y,z,p_x,p_y,p_z)$. $z_k$ is the measured quantities $(\Phi, \theta, r)$ translated to the Cartesian system $(x, y, z)$ $x_k(+)$= the state vector after applying the Kalman filter, applied at measurement (6). The propagation matrix $\Phi_k$ is the stepper written in a matrix form. It enables the propagation of the state vector between step k-1 and step k; $\theta o$ is the rms of the multiple scattering angle[2]

$$\begin{Bmatrix} x_k(-) \\ y_k(-) \\ z_k(-) \\ px_k(-) \\ py_k(-) \\ pz_k(-) \end{Bmatrix} = \left( dT * \begin{pmatrix} AA & AB \\ BA & BB \end{pmatrix} + I \right) * \begin{Bmatrix} x_{k-1}(+/-) \\ y_{k-1}(+/-) \\ z_{k-1}(+/-) \\ px_{k-1}(+/-) \\ py_{k-1}(+/-) \\ pz_{k-1}(+/-) \end{Bmatrix} ; \phi_k = dT * \begin{pmatrix} AA & AB \\ BA & BB \end{pmatrix} + I$$

$$R_0 = \begin{pmatrix} dx & 0 & 0 \\ 0 & dy & 0 \\ 0 & 0 & dz \end{pmatrix} ; \Theta_0 = (13.6 MeV / P \bullet \beta c)\sqrt{x/X0} \bullet (1 + 0.038 \bullet \ln(x/X0))$$

$$x = r \bullet \sin\Phi; y = r \bullet \cos\Phi; z = r \bullet ctg\Theta$$

The 3x3 matrices AA= (0), BA=(0); AB and BB contain the space and momentum coefficients from the stepper.
.

### 4. OUTLOOK

The Kalman filter has been implemented in the Muon code and will be tested and optimized. It is applied from the electro-magnetic calorimeter on, through the hadron calorimeter the coil and the muon detector. It accounts for multiple scattering and other stochastic processes as well as for the dE/dx. It applies a realistic propagation using information from the data. We expect further improvement in the muon identification and reconstruction. In jets we aim for a better muon separation from neighboring hadrons.

### Acknowledgments

We want to thank Fritz Dejongh for his comments about multiple scattering angles.